\documentclass{aa}

\usepackage{graphicx}
\usepackage{txfonts}
\usepackage{subcaption}
\usepackage{lscape}
\usepackage{placeins}
\usepackage{natbib}
\bibpunct{(}{)}{;}{a}{}{,}
\usepackage{hyperref}
\usepackage{bm}

\begin{document}

   \title{Determination of the horizontal velocity field in the solar atmosphere: Method validation using 3D MHD model}

   \author{Andrii Prysiazhnyi
        \and Oleksandra Baran
        }

   \institute{Astronomical Observatory of Ivan Franko National University of Lviv, Kyryla i Methodia str., 8, Lviv, 79005, Ukraine\\
   \email{[andrii.prysiazhnyi;oleksandra.baran]@lnu.edu.ua}
   }

\abstract
{}
{We present an improved version and further validation of a method originally introduced by \citet{2016KPCB...32..145S} for reconstructing horizontal velocity fields in the solar atmosphere from physical parameters typically derived from spectroscopic observations through inversion techniques. This approach relies on the continuity equation and the assumption of negligible vertical vorticity.}
{We implemented several algorithmic modifications to allow application to large spatial grids, including the compact storage of a sparse matrix of the system of linear equations. The method was tested using snapshots from the realistic 3D MHD Bifrost simulation \texttt{en024048\_hion} of the solar atmosphere, covering heights from 20 to 980~km. Horizontal velocities were reconstructed from model density and vertical velocity values. We applied a sinc filter with a Lanczos window to the reconstructed horizontal velocity maps to reduce artefacts related primarily to the use of horizontal periodic boundary conditions.}
{In the photospheric layers, the reconstructed horizontal velocity fields show a high level of agreement with the model values, with the Pearson correlation coefficient in the range 0.8--0.9. The method performs best within granules, whereas larger discrepancies occur in intergranular lanes due to complex counter-streaming flows. In the chromospheric layers, the reconstruction quality decreases significantly with height, consistent with the increasing importance of vortex motions and the breakdown of the underlying assumption.}
{The improved method provides a reliable and efficient tool for reconstructing horizontal flows in the solar photosphere. The proposed improvements make the method applicable to large observational datasets.}


\titlerunning{Determination of the horizontal velocity field in the solar atmosphere}
\authorrunning{Andrii Prysiazhnyi \& Oleksandra Baran}
\maketitle
\nolinenumbers

\section{Introduction}

Information on the full velocity field, including both vertical and horizontal components, is essential for understanding phenomena in the solar atmosphere, particularly for detailed studies of convective and wave dynamics. The component of the plasma velocity along the line of sight can be determined from spectroscopic observations via the Doppler effect. The line-of-sight velocity component corresponds mainly to the vertical velocity at the solar disc centre and contains an increasing contribution from horizontal motions towards the solar limb \citep{2010KPCB...26..117B,2011A&A...532A.110V}.

A number of dedicated methods have been developed to estimate horizontal velocities in the solar atmosphere. One of the most widely used techniques is the local correlation tracking (LCT) method, in which the cross-correlation of subregions from two successive frames of an image time series is computed. This method was proposed by \citet{1970PatRe...2..279L,1971JApMe..10..118L} to track cloud motions based on observations from geosynchronous Earth satellites and was introduced into solar physics by \citeauthor{1988ApJ...333..427N} (\citeyear{1988ApJ...333..427N}; see also \citeauthor{1986ApOpt..25..392N}~\citeyear{1986ApOpt..25..392N}). Since then, numerous variants of the LCT method have been proposed to determine horizontal velocities in the solar photosphere and chromosphere. For example, \citet{2011A&A...529A.153V} adapted the LCT algorithm for G-band observations from the Solar Optical Telescope on board Hinode, and \citet{2012AN....333..125B} modified it for continuum images from the Helioseismic and Magnetic Imager (HMI) of the Solar Dynamics Observatory (SDO). Image subregions are commonly apodised using a Gaussian window with a width comparable to the size of the tracked structures. Accordingly, the apparent horizontal velocity at each pixel can be obtained from the measured displacement by taking into account the image scale and the time interval between successive frames.

\citet{2004ApJ...610.1148W} developed an improved version of the correlation tracking method, referred to as Fourier local correlation tracking (FLCT), which computes the cross-correlation in Fourier space, thereby significantly reducing the computational cost. \citet{2008ASPC..383..373F} proposed a new variant of this method employing a two-dimensional surface-fitting procedure to locate the maximum of the cross-correlation function. This approach enables a more reliable determination of the peak position and improves the overall efficiency of the algorithm.

\citet{2013A&A...555A.136V} used synthesised continuum images based on magnetohydrodynamic (MHD) simulations to assess the reliability, accuracy, and parameter dependence of the LCT variant proposed in \citet{2011A&A...529A.153V}. The authors concluded that, while LCT reproduces the main properties of solar granulation, intrinsic horizontal velocities may be underestimated by up to a factor of three. The best agreement between the model data and the horizontal flow patterns obtained from LCT was found at the level $\log\tau_{500}=+1$, whereas a similar analysis by \citet{2014A&A...563A..93Y} yielded a maximum correlation at $\log\tau_{500}\approx0$.

\citet{2015SoPh..290.1135L} tested the performance of the LCT method on simulated convection using different apodisation windows. The authors showed that a triangular window provides the best correlation with the model data, whereas a trapezoidal window gives the poorest results. The largest differences between the model and LCT-derived data were found at granule boundaries.

Algorithms for the automated identification and tracking of individual features in solar images have also been developed. Early examples include those by \citet{1995ESASP.376b.213S,1995ESASP.376b.219S}, where a pattern-recognition approach was used to identify features in 2D observational data and track their displacements over time, enabling the reconstruction of horizontal velocity fields in the solar photosphere. A modern example of such an approach is the algorithm introduced by \citet{2025A&A...693A..71B} for tracking magnetic elements. One such feature-tracking technique is the coherent structure tracking (CST) method \citep{1999A&A...349..301R,2001A&A...377L..14R,2007A&A...471..687R,2013A&A...552A.113R}, in which the tracked features are granules, and the goal of the method is to infer large-scale surface plasma flows responsible for their motions. \citet{2001A&A...377L..14R} applied the LCT and CST methods to photospheric images obtained from realistic numerical simulations of convection and concluded that neither method is capable of reliably reproducing horizontal velocities on spatial scales smaller than 2500~km and temporal scales shorter than 30~min.

\citet{2004A&A...424..253P} proposed an original approach to detecting horizontal motions in the solar photosphere -- the so-called `balltracking' method, in which the intensity in an image sequence is treated as a dynamically evolving three-dimensional physical surface. This method achieves an accuracy comparable to LCT at a significantly lower computational cost. An improved version of the method is described in \citet{2018SpWea..16.1143A}.

The methods discussed above are purely image-processing techniques. However, horizontal velocity fields can also be inferred using physically motivated approaches, most notably those based on the induction equation that relates magnetic-field evolution to plasma motions. The theoretical foundation of such approaches was established by \citet{2002ApJ...577..501K}, and \citet{2004ApJ...610.1148W} introduced the inductive local correlation tracking (ILCT) method, enabling the reconstruction of the full velocity field under the assumption that LCT-derived horizontal flows are consistent with the vertical component of the induction equation.

Apart from methods that combine correlation tracking with corrections based on the induction equation, approaches that do not rely on LCT have also been developed. An example is the differential affine velocity estimator (DAVE) method \citep{2006ApJ...646.1358S}, which assumes that the velocity field governing the evolution between successive magnetograms can be locally represented by an affine model. The affine coefficients are determined by minimising the mismatch between the observed magnetogram differences and those predicted by the vertical component of the induction equation in continuity form. A related method, the non-linear affine velocity estimator \citep[NAVE;][]{2005ApJ...632L..53S}, allows the use of magnetograms with larger temporal separations, although at the expense of a higher computational cost compared to DAVE. \citet{2008ApJ...683.1134S} proposed a modified version of DAVE, termed the differential affine velocity estimator for vector magnetograms (DAVE4VM). This method uses vector magnetograms and, in contrast to the original DAVE method, accounts for all three components of the magnetic field vector, thereby enabling the recovery of the full velocity field. \citet{2012AAS...22020706S} introduced a further improved variant, termed the differential affine velocity estimator for vector magnetograms with Doppler velocities (DAVE4VMwDV). This approach uses observed Doppler velocities as an additional constraint for determining the horizontal velocity field. \citet{2023ApJ...955...40L} applied this enhanced method to observational data, and \citet{2025ApJ...979..139L} subsequently tested it using an MHD simulation. Earlier, a combination of vector magnetograms and Doppler velocities as input data was used by \citet{2006ApJ...636..475G}. Another approach, based on the assumption that the velocity field is consistent with the observed evolution of the magnetic field, is the minimum energy fit (MEF) method of \citet{2004ApJ...612.1181L}. In this method, among all velocity fields that satisfy the induction equation, the one that minimises the integral of the squared velocity magnitude is selected. \citet{2007ApJ...670.1434W} compared several induction equation-based methods, and \citet{2008ApJ...689..593C} extended this work by testing LCT, DAVE, and NAVE. The latter study found that NAVE best detects subpixel and non-uniform motions, whereas LCT is the fastest but exhibits difficulties in detecting non-uniform flows, while DAVE underestimates velocities on larger scales.

Methods of local helioseismology are used to infer the structure of horizontal flows in the subsurface layers of the Sun \citep[see e.g.][]{2005LRSP....2....6G,2010ARA&A..48..289G}. Several studies have compared horizontal velocity fields obtained with local helioseismology to those derived using the image-processing and induction equation-based methods described above. \citet{2007SoPh..241...27S} compared time-distance helioseismology and LCT results using observations from the Michelson Doppler Imager (MDI) of the Solar and Heliospheric Observatory (SOHO). In contrast, \citet{2007ApJ...657.1157G} used realistic simulations of solar convection, complemented by SOHO/MDI observations. In both studies, good agreement between the two approaches was found. \citet{2013SoPh..287..279L} compared photospheric horizontal velocities obtained with DAVE4VM to subsurface flows at a depth of 0.5~Mm derived from helioseismic inversions \citep{2012SoPh..275..375Z} for two active regions using SDO/HMI data. The results showed that for a simple, well-developed active region, horizontal flows in both layers exhibit a similar pattern. \citet{2013ApJ...771...32S} used SDO/HMI data to compare CST and local helioseismology results, demonstrating good agreement between the two methods. Based on these findings, \citet{2018A&A...611A..92R} extended the analysis using a much longer time series allowing the derivation of global solar properties, such as differential rotation. Agreement between helioseismology results and LCT data was also reported by \citet{2017A&A...606A..28L} in a study of inflows to active regions.

Over the past decade, machine learning methods have been increasingly applied in solar physics. In this context, \citet{2017A&A...604A..11A} introduced the convolutional neural network DeepVel trained on MHD-simulated velocity fields to infer horizontal velocities at multiple atmospheric heights from pairs of consecutive intensity images; comparisons with LCT showed similar results and a substantial gain in computational speed. Its performance depends on the spatial and temporal resolution as well as the physical model used to train the network. Versions of DeepVel have been trained for both quiet-Sun and active-region observations. \citet{2018SoPh..293...57T} compared DeepVel with other horizontal-velocity determination methods (LCT, FLCT, and CST) using synthetic intensitygrams. The study showed that DeepVel reproduces horizontal motions most accurately in quiet regions at granular and subgranular scales, whereas FLCT performs better at supergranular scales. Nevertheless, the authors argued that if data with the corresponding spatial resolution are used to train DeepVel, this network will outperform classical tracking methods even at supergranular scales. \citet{2020FrASS...7...25T} developed a new version of DeepVel using a U-Net architecture, termed DeepVelU, which demonstrated a substantial increase in correlation between model and reconstructed velocities compared to the original DeepVel. \citet{2022A&A...658A.142I} proposed a new convolutional neural network-based method to infer horizontal velocity fields from spatio-temporal variations in intensity and vertical velocity trained on MHD simulations of turbulent convection. The accuracy of the reconstructed horizontal velocities was evaluated at different spatial scales using coherence spectra, and a comparison with DeepVel showed the advantages of the new method. \citet{2025A&A...698A.263L} developed a series of machine-learning models to infer horizontal velocities in the photosphere. These models are based on a shallow U-Net architecture and were trained using intensity, the vertical magnetic field component, and horizontal velocities from MHD simulations. Comparison of this approach, termed SUVEL, with FLCT and DeepVel demonstrated its superiority over the other methods. The current version of SUVEL, however, is only applicable to horizontal-velocity inference in quiet-Sun regions of the photosphere.

\citet{2016KPCB...32..145S} proposed another approach that accounts for the physics of the investigated region when determining horizontal velocities. This method reproduces the field of horizontal velocities on granular and subgranular scales based on observations of the solar disc centre with high spatial and temporal resolution. In this method, semi-empirical models of the solar atmosphere are constructed by solving the inverse radiative transfer problem, and hydrodynamic equations are then used to infer the horizontal velocity components from the physical parameters of the model. Tests using snapshots from a 3D hydrodynamic simulation demonstrated that this method can reliably reproduce the horizontal velocity field at different heights in the solar photosphere, although the reconstruction quality decreases in the upper photospheric layers. \citet{2019KPCB...35..231S} applied this method to determine the horizontal velocity field in the vicinity of small-scale high-speed plasma flows (photospheric jets) in order to infer their structure and formation mechanism.

A somewhat similar method for retrieving horizontal velocity fields at different heights in the solar atmosphere was proposed by \citet{2026A&A...707A.306V}. In this approach, the horizontal velocities are determined from the inversion results using the ideal MHD induction equation. Unlike the method of \citet{2016KPCB...32..145S}, this approach requires spectropolarimetric observations to infer the magnetic field components and their temporal derivatives through inversion. Another important difference is that the horizontal velocity field cannot be determined independently at a selected height. Instead, the method solves a single large system of linear equations to retrieve the horizontal velocities simultaneously at all considered heights. Furthermore, in its current implementation, the method has only been demonstrated for reconstructing the horizontal velocity field in a single vertical plane.

We have developed an improved version of the horizontal velocity calculation method originally proposed by \citet{2016KPCB...32..145S} specifically optimised for application to large spatial grids. Here, we report on how we tested it using a publicly available 3D MHD simulation of the solar atmosphere.

The remainder of this paper is organised as follows. The method is described in Sect.~\ref{sec:methodanddata} along with the data used for its testing. The results of our analysis and their discussion are provided in Sect.~\ref{sec:results}. Finally, the conclusions are presented in Sect.~\ref{sec:conclusions}.

\section{Method and data}
\label{sec:methodanddata}

Both the method proposed by \citet{2016KPCB...32..145S} and its improved version presented in this work rely on constructing a 3D semi-empirical atmospheric model by solving the inverse radiative transfer problem. The gas density and vertical velocity provided by the model (expressed on a geometrical height scale) are then used in a separate system of equations to recover the unknown horizontal velocity components. In this way, the proposed approach complements the inversion results by using them as input parameters. In principle, any inversion code capable of producing a 3D atmospheric model can be used for this purpose, whether it operates in the 1.5D approximation (where the model is inferred independently at each spatial pixel) or accounts for 2D or full 3D radiative transfer, and whether it assumes local thermodynamic equilibrium (LTE) or includes non-LTE effects. The reliability of the recovered horizontal velocity fields depends directly on the quality of the atmospheric model inferred from the inversion but does not affect the validity of the method itself.

To determine the horizontal velocity components, $u_x$ and $u_y$, at each spatial point, we used the hydrodynamic continuity equation:
\begin{equation*}
	\frac{\partial\varrho}{\partial t}+\nabla\cdot(\varrho\bm{u})=0,
\end{equation*}
where $\varrho$ is the mass density and $\bm{u}$ is the flow velocity. To obtain a closed system, this equation must be supplemented by an additional relation involving the same unknowns. In the present version of the method, the second equation is obtained under the assumption that the vertical component of the vorticity, $\omega_z$, is zero. Although this assumption is not directly motivated by physical considerations, it provides the simplest mathematical formulation for the required supplementary relation involving $u_x$ and $u_y$:
\begin{equation*}
	\omega_z=\big(\nabla\times\bm{u}\big)_z=0\Rightarrow\frac{\partial u_y}{\partial x}-\frac{\partial u_x}{\partial y}=0.
\end{equation*}
For a horizontal spatial grid with $N$ points in each direction, this formulation results in a system of $2N^2$ linear differential equations with the same number of unknown horizontal velocity components. Using a finite-difference approximation for the derivatives, this system can be converted to a system of linear algebraic equations. We adopted a fourth-order centred-difference scheme based on a five-point stencil. We let $i$ denote the grid-point index along a given spatial direction, and then the derivative of a quantity $f$ was approximated by
\begin{equation*}
	\left(\frac{\partial f}{\partial x}\right)_i\approx\frac{f_{i-2}-8f_{i-1}+8f_{i+1}-f_{i+2}}{12\Delta},
\end{equation*}
where $\Delta$ is the spatial grid spacing. An analogous fourth-order centred-difference approximation was used to compute the time derivative. 

The system was solved using the iterative steepest descent method, which requires the coefficient matrix $\mathbf{A}$ to be symmetric. To ensure the symmetry required by the iterative procedure, both sides of the equation were multiplied on the left by $\mathbf{A}^{\rm\!T}$, the transpose of the coefficient matrix.

For relatively large values of $N$, the coefficient matrix becomes large and requires a substantial amount of random-access memory (RAM) for storage. In addition, matrix multiplications involving such large matrices are computationally expensive. To test the method, \citet{2016KPCB...32..145S} employed the 3D hydrodynamic model of \citet{2000A&A...359..669A} with a spatial grid of $50\times50$ points. Later, \citet{2019KPCB...35..231S} computed the horizontal velocity field in the vicinity of a photospheric jet on a $100\times100$ grid. In these studies, the entire coefficient matrix $\mathbf{A}$ was stored in RAM as a regular two-dimensional array. For relatively small spatial grids, as in the studies mentioned above, this approach is acceptable. However, for the larger grid employed in the present work, it is not practical since the vast majority of the matrix elements are zero.

To store the matrix $\mathbf{A}$ more compactly, we exploited its sparsity: It contains only 26 diagonals with non-zero elements. The sparsity of the matrix not only enables compact storage in memory but also allows the multiplication for computing $\mathbf{A}^{\rm\!T}\!\mathbf{A}$ to be carried out much more efficiently. Computing $\mathbf{A}^{\rm\!T}\!\mathbf{A}$ reduces to multiplying the columns of $\mathbf{A}$, which yields a non-zero result only when at least one row contains non-zero elements in both columns. The resulting matrix $\mathbf{A}^{\rm\!T}\!\mathbf{A}$ contains 143 diagonals, whose positions are determined by the difference in column indices of $\mathbf{A}$. The diagonals of the resulting matrix are partitioned into segments, each corresponding to contributions from specific pairs of diagonals of $\mathbf{A}$. Using information on the boundaries of these segments, together with the indices of the diagonal pairs, the matrix multiplication can be performed efficiently. For compact storage of $\mathbf{A}^{\rm\!T}\!\mathbf{A}$, it is sufficient to retain only the elements of the main diagonal and all upper (or lower) diagonals. The total number of elements contained in all non-zero diagonals of the matrix $\mathbf{A}$ is $35N^2-4N$, whereas the number of elements of $\mathbf{A}^{\rm\!T}\!\mathbf{A}$ that need to be stored in memory is $79N^2-6N+7$. Figure~\ref{fig:matrices} schematically illustrates the structures of these matrices. The total number of elements in each of the full matrices is $4N^4$.

For a spatial grid of $100\times100$ pixels, storing the non-zero elements of $\mathbf{A}$ in a compact form using double precision format requires 2.67~MB of RAM, while $\mathbf{A}^{\rm\!T}\!\mathbf{A}$ requires 6.02~MB; in contrast, the full matrix would occupy 2.98~GB. Thus, the use of compact representations for these matrices reduces the memory requirements by three orders of magnitude compared to the full matrix representation.

In practice, however, we adopt a less memory-optimal but computationally more efficient compact storage scheme for the matrix $\mathbf{A}$. The matrix is stored as a regular two-dimensional array that offers better computational performance than other types of data collections. Each subarray has the same length and contains the elements of a particular matrix row corresponding to the 26 diagonals. This implies that memory is allocated even for elements corresponding to diagonals that do not intersect a given row; these elements are never used in the computations. As a result, the total number of stored elements is $52N^2$ (3.97~MB for the $100\times100$ grid), which is still more than two orders of magnitude smaller than in the full matrix representation.

Knowing the offsets of the diagonals containing non-zero elements in the matrix $\mathbf{A}$, one can determine the number of addition and multiplication operations required to compute the product $\mathbf{A}^{\rm\!T}\!\mathbf{A}$. This makes it possible to estimate the reduction in computational cost compared to standard matrix multiplication. For the standard multiplication of two square matrices of size $n$, the numbers of addition and multiplication operations are $n^3-n^2$ and $n^3$, respectively. For the matrices considered here, $n=2N^2$, resulting in $8N^6-4N^4$ addition operations and $8N^6$ multiplication operations. For the improved multiplication proposed here, these numbers are reduced to $245N^2-62N+3$ and $324N^2-68N+10$, respectively. For a spatial grid with $N=100$, the proposed scheme therefore requires only $4\times10^{-5}$\% of the addition operations and $3\times10^{-5}$\% of the multiplication operations compared to the standard approach.

For the multiplication of $\mathbf{A}^{\rm\!T}$ by the vector $\mathbf{b}$, the standard approach requires $4N^4-2N^2$ addition operations and $4N^4$ multiplication operations, whereas the improved approach requires only $33N^2-4N$ and $35N^2-4N$ operations, respectively. Thus, for a $100\times100$ grid, the improved matrix--vector multiplication requires only 0.09\% of the addition operations and 0.08\% of the multiplication operations compared to the standard approach.

For the iterative procedure, the standard approach requires $8N^4$ addition operations and $8N^4+6N^2$ multiplication operations per iteration, whereas the accelerated approach reduces these numbers to $312N^2-24N+28$ and $318N^2-24N+28$, respectively. For the example considered above, this corresponds to a reduction to 0.39\% (additions) and 0.40\% (multiplications).

To test the method for calculating the horizontal velocity field, we used a publicly available realistic 3D MHD simulation \texttt{en024048\_hion} of the solar atmosphere\footnote{\url{http://sdc.uio.no/search/simulations}}, obtained with the Bifrost code \citep{2011A&A...531A.154G}. The simulated quiet-Sun region corresponds to an enhanced network area. The simulation covers heights from the upper part of the convection zone to the lower corona. The size of the computational box is 24~Mm in each of the horizontal directions and 16.8~Mm in the vertical direction. The horizontal resolution is 48~km, while the vertical resolution varies from 19~km in the lower atmosphere to 100~km at the upper boundary. The spatial grid comprises $504\times504$ points in the horizontal planes and 496 vertical levels. The zero point of solar time is defined as the moment when the chromosphere and corona were added to the initial model consisting only of the upper part of the convection zone and the photosphere. The first snapshot was taken at solar time $t=3850$~s (snapshot 385), and subsequent snapshots were recorded every 10~s. A detailed description of this simulation is provided by \citet{2016A&A...585A...4C}.

\section{Results and discussion}
\label{sec:results}

In this work, we computed the horizontal velocities for each snapshot of the \texttt{en024048\_hion} simulation from 391 to 530, resulting in 140 time instances. To apply the method of \citet{2016KPCB...32..145S}, a uniform vertical spacing in each 3D model is required. Therefore, we interpolated the original simulation data at each spatial grid point using cubic splines, producing a set of 3D models covering the height range from 0 to 1000~km with a vertical spacing of 10~km. The horizontal velocity fields were calculated in the height range 20--980~km, because the finite-difference approximation employed for derivative evaluation requires values at two adjacent grid points above and below each height level. To reduce the code execution time, we computed horizontal velocities only for the central part of the spatial grid ($252\times252$ points).

It should be noted that, when determining the horizontal velocity field from semi-empirical models obtained by inversion of observational data, the effective spatial resolution in the vertical direction is expected to be substantially lower than that adopted here. In inversion procedures, the physical parameters are modified only at a relatively small number of height points (nodes), while the values at intermediate points are obtained by interpolation and therefore do not provide additional independent information. In the present work, however, we did not perform inversions, but used the density and vertical velocity directly from the MHD model, in which the vertical resolution in the photospheric layers is 19~km \citep{2016A&A...585A...4C}. This allowed us to use more detailed height stratifications than are typically available in semi-empirical models derived from inversions; therefore the use of a uniform grid with a spacing comparable to this smallest model step was justified.

We performed additional calculations of the horizontal velocity fields for a single snapshot (snapshot 391) at 21 height levels in the range 200--400~km (this range corresponds to the high correlation between the model and reconstructed horizontal velocities; see Fig.~\ref{fig:corrcoeff} below) in order to examine how the vertical spacing used in the finite-difference evaluation of derivatives affects the retrieved horizontal velocities. These calculations show that nearly identical results are obtained for vertical grid spacings between 10 and 100~km.

The availability of reference (model) values of horizontal velocities enables the introduction of a numerical criterion that characterises the difference between the reproduced and model velocity fields for the same area. This allowed us to evaluate the reconstruction quality at each iteration, i.e. the proximity to the model values of horizontal velocities. As a criterion we used the sum of absolute differences between the original and reproduced $x$- and $y$-components of velocity for all spatial grid points. After the iteration corresponding to the minimum distance, the quality of the solution deteriorates, i.e. the iterative procedure produces velocity fields that increasingly deviate from the reference values.

\begin{figure}[h!]
	\centering
	\includegraphics[width=8.8cm]{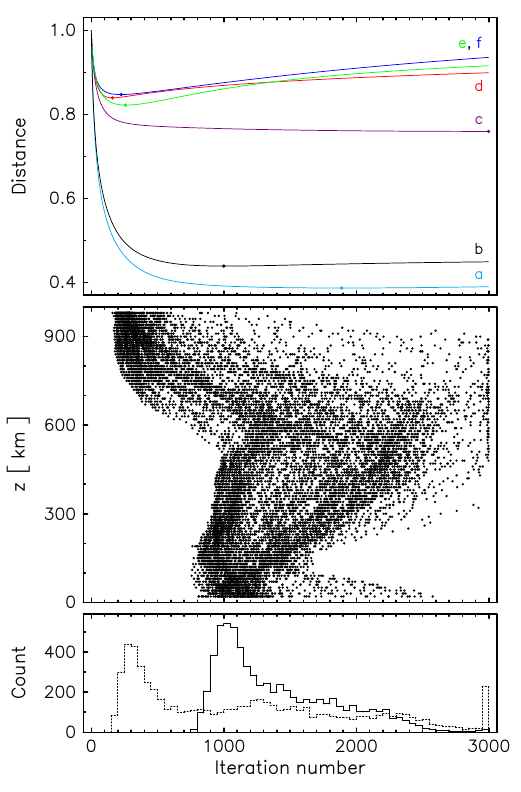}
	\caption{Characteristics of the iterative procedure. \textit{Top panel}: Selected curves showing the distance between the original (model) and calculated horizontal velocity fields, normalised to the first-iteration distance. The curves correspond to specific cases: the largest (\textit{a}) and smallest (\textit{f}) differences between the first-iteration distance and the minimum distance, the largest (\textit{c}) and smallest (\textit{d}) iteration numbers at which the minimum is reached, the steepest increase after the minimum (\textit{e}), and a representative curve with a minimum near iteration 1000 (\textit{b}). \textit{Middle panel}: Scatter plot of the iteration numbers corresponding to the minima for different heights. \textit{Bottom panel}: Histograms of the minimum positions for heights below 500~km (solid line) and above 500~km (dashed line).}
	\label{fig:distances}
\end{figure}

The top panel of Fig.~\ref{fig:distances} shows several representative examples of the criterion characterising the distance between the reference and calculated horizontal velocity fields as a function of iteration number. Each curve is normalised to its value at the first iteration. The selected cases include the largest and smallest differences between the first-iteration distance and the minimum distance, the largest and smallest iteration numbers at which the minimum distance occurs, the steepest increase in the distance after the minimum, and a representative case of the iterative procedure. In all calculations, we used 3000 iterations, which proved sufficient in nearly all cases. 

The middle panel of Fig.~\ref{fig:distances} shows a scatter plot illustrating the distribution of the iteration numbers at which the minimum distance is reached for different heights. In the lower atmospheric layers, the optimal solution is typically reached after 900--1100 iterations, while the number of iterations required to reach the optimum generally increases with height. However, above approximately 500~km, the opposite trend is observed, with the optimal iteration number decreasing. The separation is even more pronounced in the bottom panel of Fig.~\ref{fig:distances}, which presents the corresponding histograms for heights below and above 500~km. As discussed later and shown in Fig.~\ref{fig:corrcoeff}, starting from approximately the same heights, the correlation coefficient between the model and reconstructed velocity fields begins to decrease with height. In some cases, for approximately 1\% of the iterative runs, the best agreement with the model data is reached at the final iteration, indicating that a larger number of iterations may be required. 

When dealing with real observational data, a convergence criterion based on the relative or absolute difference between the solutions obtained in several consecutive iterations should be employed to terminate the iterative procedure once it stabilises, since further iterations only lead to negligible changes. Alternatively, a fixed number of iterations may be adopted.

\begin{figure}[h!]
	\centering
	\includegraphics[width=8.8cm]{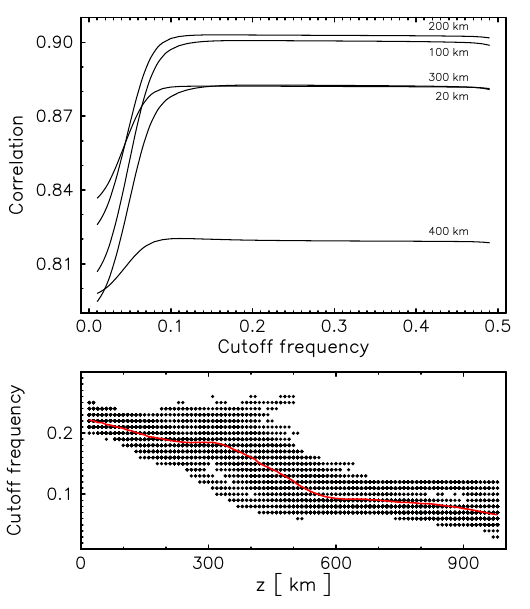}
	\caption{Determination of the optimal cutoff frequency. \textit{Top panel}: Pearson correlation coefficient between the original and reconstructed horizontal velocity fields after filtering as a function of cutoff frequency for selected atmospheric heights (snapshot 391). \textit{Bottom panel}: Scatter plot showing the optimal (maximum-correlation) cutoff frequency for each height and snapshot. The solid red curve indicates the height dependence of this parameter averaged over all snapshots.}
	\label{fig:optfreq}
\end{figure}

\begin{figure}[h!]
	\centering
	\includegraphics[width=8.8cm]{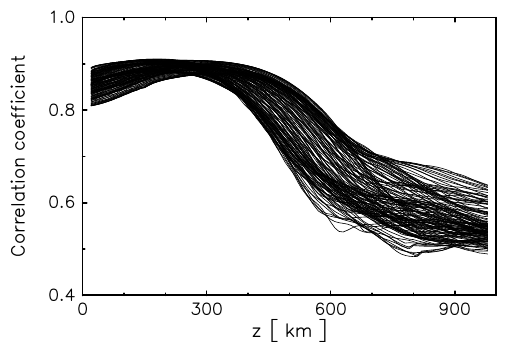}
	\caption{Pearson correlation coefficient between the original horizontal velocities and the calculated values after filtering as a function of height for snapshots 391--530.}
	\label{fig:corrcoeff}
\end{figure}

\begin{figure}[h!]
	\centering
	\includegraphics[width=8.8cm]{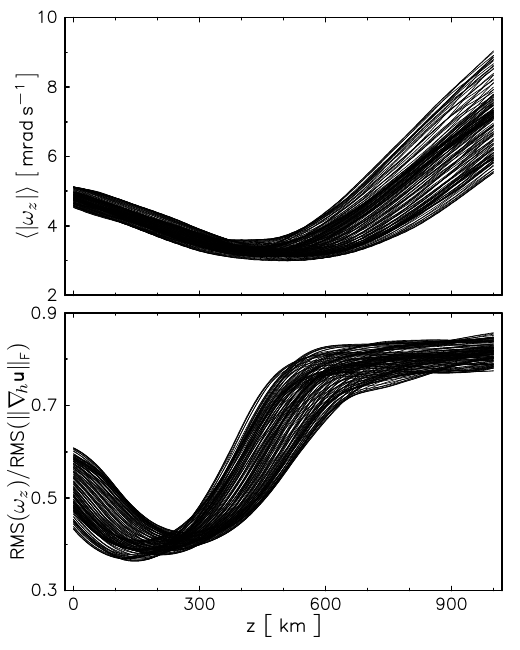}
	\caption{Vertical vorticity as a function of height. \textit{Top panel}: Absolute value of the vertical vorticity, $|\omega_z|$, averaged over the analysed area for snapshots 391--530. \textit{Bottom panel}: Ratio of the RMS of $\omega_z$ to the RMS of the Frobenius norm of the horizontal velocity gradient tensor.}
	\label{fig:omega}
\end{figure}

\begin{figure}[h!]
	\centering
	\includegraphics[width=8.8cm]{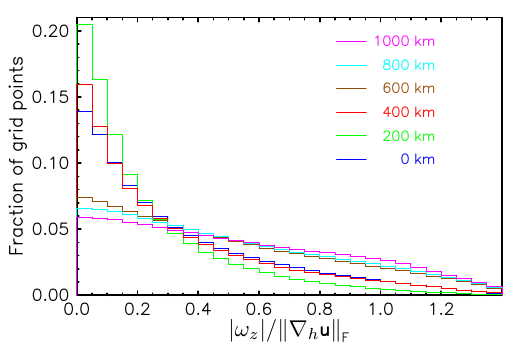}
	\caption{Histograms of the normalised $|\omega_z|$ distribution for selected heights, including all snapshots (391--530). The bin width is 0.05. The fraction shown on the $y$-axis is defined relative to the total number of grid points at each height, including all snapshots.}
	\label{fig:omegahistograms}
\end{figure}

\begin{figure*}[h!]
	\centering
	\includegraphics[width=17cm]{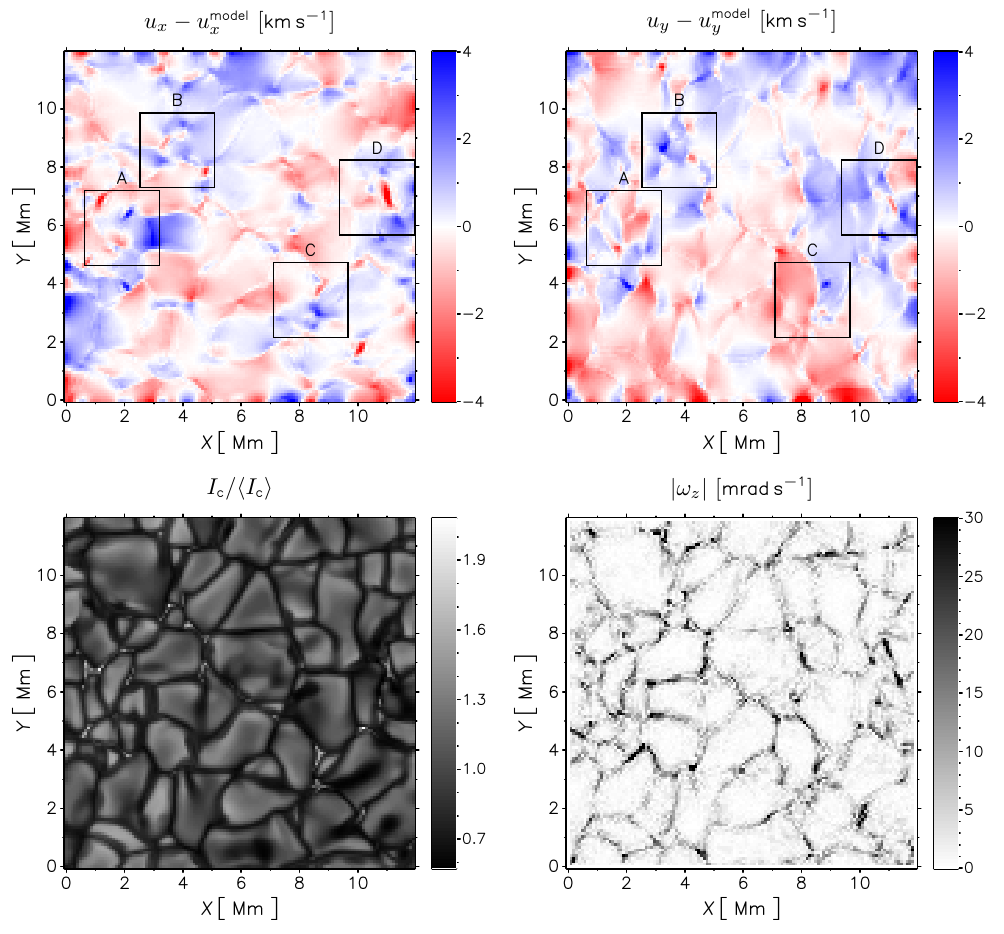}
	\caption{Spatial relationship between reconstruction quality and vertical vorticity. \textit{Upper left panel}: Difference between the calculated values of the $x$-component of the horizontal velocity after filtering and the model data for snapshot~391 at a height of 160~km. \textit{Upper right panel}: Same but for the $y$-component. \textit{Lower left panel}: Normalised continuum intensity near the \ion{Mg}{ii} line (4482~\AA). \textit{Lower right panel}: Surface map of the absolute value of the vertical vorticity, $|\omega_z|$, for snapshot~391 at 160~km. Black squares indicate regions shown in more detail in Fig.~\ref{fig:vectors}.}
	\label{fig:diffandomega}
\end{figure*}

To assess the computational speed-up provided by our improved matrix multiplication scheme, we computed the horizontal velocity fields using both the standard matrix multiplication and the proposed approach for the same height range and snapshot as in the calculations of the effect of the vertical step. In the case of standard matrix multiplication, the most computationally expensive step is the evaluation of $\mathbf{A}^{\rm\!T}\!\mathbf{A}$. For the accelerated scheme, the time required for this operation is drastically reduced and is, on average, approximately 0.007\% of that in the standard case over the considered height range. For the multiplication of $\mathbf{A}^{\rm\!T}$ by the right-hand-side vector $\mathbf{b}$, the computation time is likewise approximately 1.5\% of that in the standard case; however, the contribution of this operation to the total computational cost is negligible. For the iterative process, the speed-up is significantly smaller. Nevertheless, the use of accelerated matrix--vector multiplication reduces the time required for the iterations to about 15\% of that in the original approach. The speed-up obtained in these calculations is smaller than that described in Sect.~\ref{sec:methodanddata}, since it is not determined solely by the difference in the number of addition and multiplication operations, but also depends on the implementation details (for example, the additional summation operations required to compute offsets of matrix and vector elements in the underlying data arrays).

The obtained surface distributions of horizontal velocities contain artefacts that manifest primarily as patterns of horizontal and vertical stripes. We attribute these artefacts to periodic horizontal boundary conditions. In addition, localised regions exist, mainly in intergranular lanes, where the method is less effective, resulting in pronounced, abrupt changes in the horizontal velocity, both in magnitude and direction. To reduce the artefacts, we applied a sinc filter with a Lanczos window, first along one horizontal coordinate (independently for each row) and then, using the same filter parameters, along the other coordinate (independently for each column). It should be noted that this filtering procedure was not intended to substantially improve the reconstruction quality, but only to remove artefacts that are visually prominent in the velocity fields.

\begin{figure*}[h!]
	\centering
	\includegraphics[width=17cm]{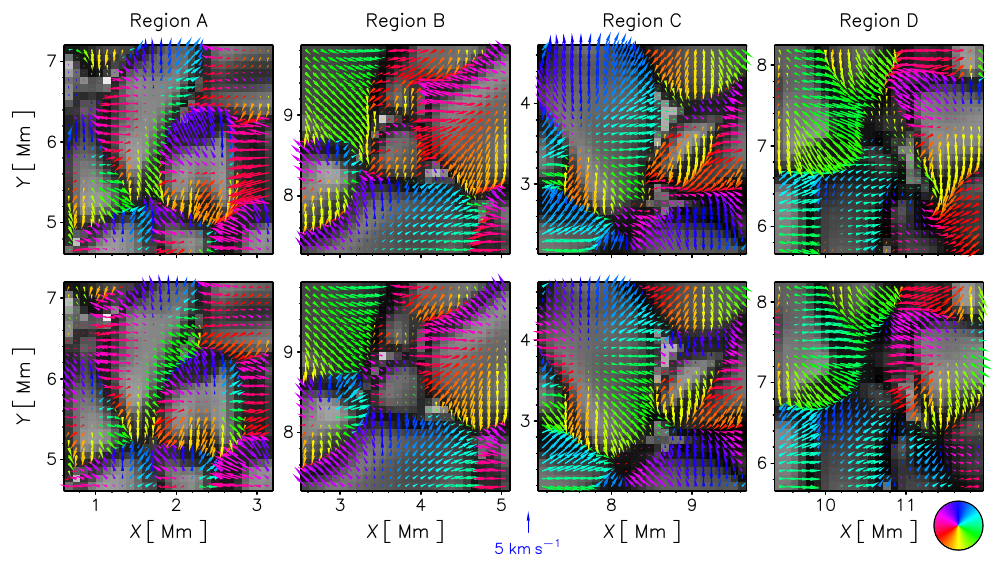}
	\caption{Model (\textit{top row}) and calculated (\textit{bottom row}) horizontal velocity fields shown by coloured arrows and overlaid on the radiation intensity map at 4482~\AA\ for the selected regions. The correspondence between the colours of the vectors and their directions is indicated by the colour wheel. The scale is shown at the bottom by a vector representing a horizontal velocity of 5~${\rm km\,s^{-1}}$.}
	\label{fig:vectors}
\end{figure*}

A crucial aspect of the filtering procedure is the selection of an appropriate filter cutoff frequency that removes artefacts without excessively smoothing the surface distributions. To determine this frequency, we filtered the obtained horizontal velocity fields for all snapshots and heights using a range of cutoff frequencies and calculated the Pearson correlation coefficient between the original data from the MHD model and the reproduced horizontal velocities after filtering. The top panel of Fig.~\ref{fig:optfreq} shows examples of the correlation coefficients as a function of cutoff frequency at several atmospheric heights for a selected snapshot. In all cases, the correlation increases with increasing cutoff frequency up to a cutoff frequency of approximately 0.1, reaches a plateau extending to about 0.45, and then exhibits a slight decrease. The highest correlation is obtained, on average, near a cutoff frequency of 0.14. However, the variations within the plateau are negligible, indicating that any value within this range is suitable for filtering. The scatter plot in the bottom panel of Fig.~\ref{fig:optfreq} shows the optimal (maximum-correlation) cutoff frequency as a function of height for individual snapshots. Since it varies systematically with height, we adopted, for each height level, the cutoff frequency averaged over all snapshots. The filter effectively suppresses the artefacts while preserving the overall structure of the velocity field (see Figs.~\ref{fig:artefacts1} and~\ref{fig:artefacts2}).

We calculated the Pearson correlation coefficient between the reproduced horizontal velocities after filtering and the corresponding reference values from the MHD model for heights between 20~km and 980~km. The resulting curves for selected heights and all studied snapshots are shown in Fig.~\ref{fig:corrcoeff}. For all snapshots, the correlation coefficient is in the range 0.8--0.9 for heights up to approximately 400~km, but decreases with height, reaching about 0.5--0.6 near the upper boundary of the analysed height range. The correlation coefficient calculated from the unfiltered data differs only slightly from that shown in Fig.~\ref{fig:corrcoeff}.

Since we used the density and vertical velocity directly from the model, our results are not affected by errors associated with the inversion procedure. Consequently, one of the main sources of error is expected to be the approximation of a negligible vertical component of vorticity, which is inherent to the method. To assess the impact of this approximation on the results, we calculated, for each height level, the mean value of $|\omega_z|$ over the entire sequence of snapshots (top panel of Fig.~\ref{fig:omega}). The magnitude of $\langle|\omega_z|\rangle$ decreases with height, reaching a minimum at $z=400-500$~km, and then increases to values even higher than in the lower atmosphere.

Additionally, we compared the vertical vorticity with the Frobenius norm of the horizontal velocity gradient tensor, which characterises all velocity gradients at a given point:
\begin{equation*}
	\|\nabla_{\!h}\mathbf{u}\|_{\rm F}^{}=\sqrt{\left(\frac{\partial u_x}{\partial x}\right)^2+\left(\frac{\partial u_x}{\partial y}\right)^2+\left(\frac{\partial u_y}{\partial x}\right)^2+\left(\frac{\partial u_y}{\partial y}\right)^2}.
\end{equation*}
We computed the ratio of root-mean-square (RMS) values ${\rm RMS}(\omega_z)/{\rm RMS}(\|\nabla_{\!h}\mathbf{u}\|_{\rm F}^{})$ for all height levels (Fig.~\ref{fig:omega}, bottom panel). In contrast to $|\omega_z|$, the RMS ratio exhibits a different height dependence: It reaches a minimum in the lower layers (200--300~km), increases up to a height of approximately 600~km, and then plateaus in the higher layers. This indicates that vortex motions contribute relatively less to the total horizontal velocity gradients in the photosphere, while their importance increases towards the chromospheric layers.

To further quantify this, we computed the normalised quantity $|\omega_z|/\|\nabla_{\!h}\mathbf{u}\|_{\rm F}^{}$. The distributions of this quantity at several height levels are shown in Fig.~\ref{fig:omegahistograms}. The fraction of grid points in each histogram was calculated relative to the total number of grid points at that height over all snapshots. At all heights, the peak occurs in the first bin. The histograms reveal a steeper decline with increasing $|\omega_z|$ in the photospheric layers compared to higher layers. The higher fractions of grid points corresponding to larger $|\omega_z|$ in the chromospheric layers indicate that vortex motions become more significant relative to the total horizontal velocity gradients. This behaviour suggests that the approximation employed in the method becomes less accurate in the upper part of the analysed height range.

\begin{figure*}[h!]
	\centering
	\includegraphics[width=17cm]{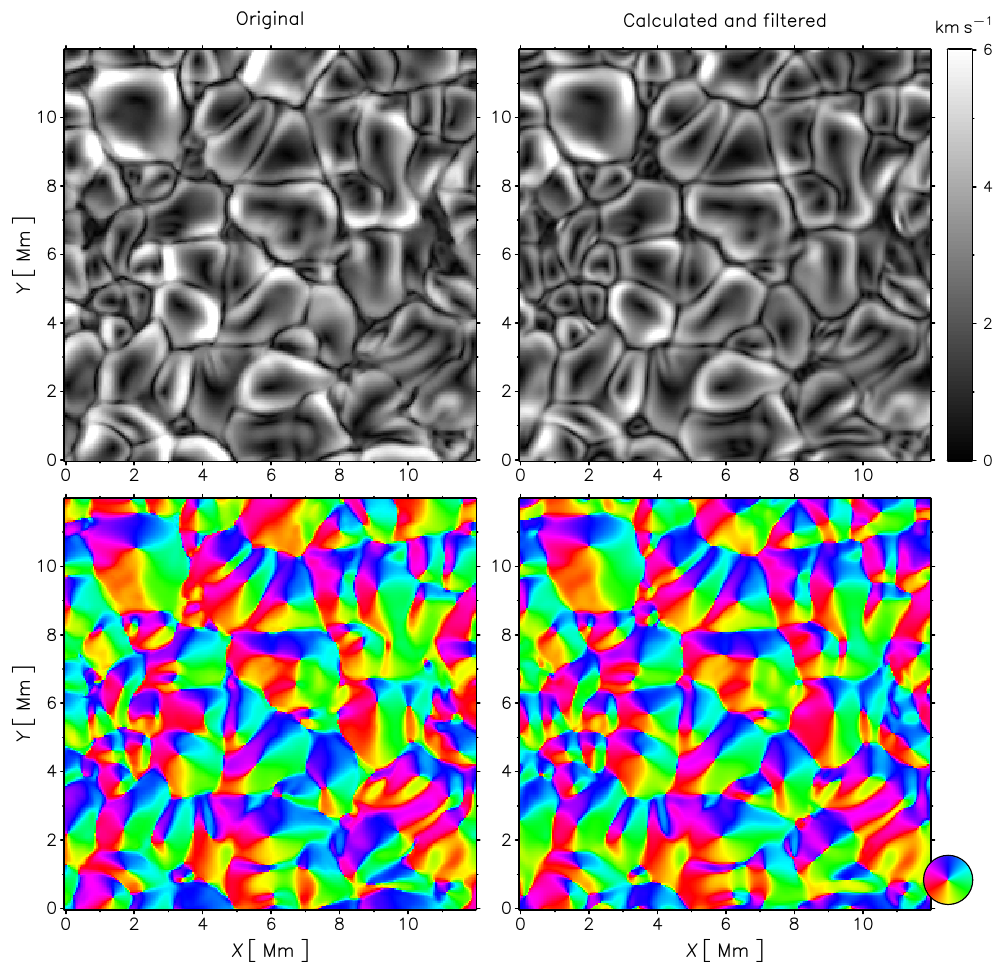}
	\caption{Model (\textit{left panels}) and calculated (\textit{right panels}) horizontal velocity fields for snapshot~391 at 160~km over the entire analysed area. The top panels show the absolute values of the horizontal velocity, and the bottom panels show the velocity directions represented by colour.}
	\label{fig:originalandfiltered}
\end{figure*}

\begin{figure}[h!]
	\centering
	\includegraphics[width=8.8cm]{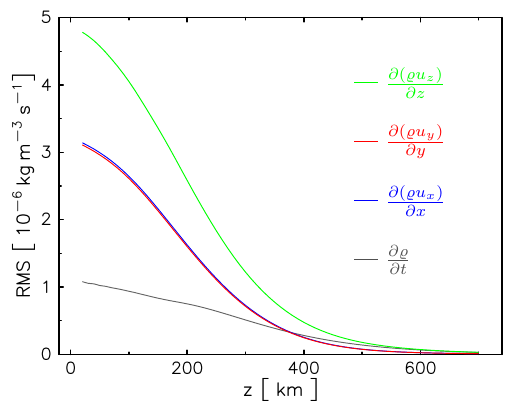}
	\caption{Height dependence of the mean RMS values (averaged over the full sequence of analysed snapshots) of the four terms entering the continuity equation: $\partial \varrho/\partial t$, $\partial (\varrho u_x)/\partial x$, $\partial (\varrho u_y)/\partial y$, and $\partial (\varrho u_z)/\partial z$.}
	\label{fig:RMSofconteqterms}
\end{figure}

To compare the distribution of $|\omega_z|$ with the granular pattern in the simulated area, we used the results of radiative transfer calculations provided with the atmospheric models. These calculations include spectral line profiles of ionised magnesium. We selected the \ion{Mg}{ii} 4481~\AA\ line. The lower left panel of Fig.~\ref{fig:diffandomega} shows the granulation pattern as seen in the continuum near this spectral line. The lower right panel shows the surface distribution of $|\omega_z|$ at a height of 160~km (snapshot 391). This height corresponds to the maximum value of the correlation coefficient between the model and reconstructed horizontal velocities for this snapshot. The highest values of $|\omega_z|$ correspond to intergranular lanes. This suggests that the approximation underlying the method allows a reliable determination of horizontal velocities within granules at lower atmospheric heights, whereas the reproduction is less accurate in intergranular lanes and in the chromospheric layers. We compared the model and reproduced values of both components of horizontal velocity for the central part of the simulated domain (upper panels of Fig.~\ref{fig:diffandomega}). The corresponding surface maps of the differences show that the largest discrepancies occur predominantly in the intergranular lanes.

We analysed the differences between the original and reproduced values of horizontal velocities for four selected regions in the central part of the simulated domain where significant discrepancies, primarily in intergranular lanes, are observed. These regions are shown in more detail in Fig.~\ref{fig:vectors}. Horizontal flows converge towards intergranular lanes in these regions. As a result, complex interactions of counter-streaming plasma flows arise, explaining the degradation in reconstruction quality. The method is based on the simplifying assumption of negligible vertical vorticity and therefore is less capable of reproducing such complex interactions on small spatial scales. However, the reconstruction quality within the surrounding granules remains satisfactory, with a sharp transition to reduced accuracy in the intergranular lanes.

Figure~\ref{fig:originalandfiltered} shows the original horizontal velocity field and the reconstructed values after filtering. The two horizontal velocity maps show differences, caused, as discussed above, by the reduced reconstruction quality of the method in regions with complex small-scale interactions of convective flows in magnetised plasma. Nevertheless, the overall quality of the reconstructed horizontal velocity field remains high.

The relative importance of the individual terms in the continuity equation determines which physical processes primarily govern the reconstruction quality. To quantify this, we evaluated the RMS values of the temporal density derivative and the three contributions of the mass-flux divergence as functions of height. Figure~\ref{fig:RMSofconteqterms} shows the RMS values of the continuity equation terms for each horizontal layer, averaged over the full sequence of snapshots. The vertical term of the mass-flux divergence provides the dominant contribution throughout most of the considered height range, while the horizontal divergence terms are intermediate and the temporal density derivative is systematically the smallest. Thus, the reconstruction is influenced primarily by the mass-flux divergence terms, with a secondary contribution from the temporal density derivative.

The results presented in this paper show that the method proposed by \citet{2016KPCB...32..145S} and further improved and tested in the present work can be used to determine the horizontal velocity field in the solar photosphere. Tests with a realistic MHD model show that the method provides significantly less reliable estimates in the chromosphere than in the photosphere.

The method successfully reproduces the dominant spatial structures of the velocity field from the MHD model. However, ensuring the robustness of the results requires careful assessment of the numerical stability of the algorithm. The potential application of preconditioning and regularisation should be considered to enhance the reliability of the results.

While the horizontal derivatives in this study are computed on a consistent geometrical height ($z$) grid, it is important to consider the implications when applying the method to observational data. In typical spectropolarimetric inversions, pixels are treated independently, which can introduce high-frequency numerical noise in spatial derivatives, as small pixel-to-pixel inconsistencies in the inferred atmospheric parameters or in the $\tau$-to-$z$ conversion can be amplified by differentiation. Because the present analysis uses the direct output of an MHD simulation, this source of uncertainty is avoided here. For future applications to observational data, however, additional spatial regularisation of the inferred atmospheric structure will likely be required for numerical stability when evaluating horizontal derivatives. A full 3D inversion \citep[see e.g.][]{2013A&A...557A.143S} would provide a more self-consistent treatment, but remains computationally expensive for routine applications.

An alternative to the finite-difference approximation for computing derivatives is the application of forward and inverse two-dimensional discrete Fourier transforms (DFT). This approach avoids the need to solve large systems of equations and can substantially reduce the computational cost of reconstructing the horizontal velocity field. We implemented this algorithm and obtained preliminary results for a single snapshot of the MHD simulation (see Appendix~\ref{app:Fourier}).

Using vertical velocity and mass density directly from the MHD model as input data allows assessment of the intrinsic uncertainties of the method and partly explains the high level of agreement between the calculated and model velocities. A more realistic test, similar to that in \citet{2016KPCB...32..145S}, would require spectral synthesis followed by inversion and subsequent application of the method to the inversion results. Such a test would allow evaluation of the combined impact of errors intrinsic to the method, those introduced by the inversion, and those associated with the conversion from optical depth to geometrical height.

Another important aspect is the effect of noise in observational data, which could be assessed by adding random noise to the synthesised profiles, with a prescribed amplitude comparable to that of real observational data. These noise-contaminated profiles would then be used as input for the inversion procedure, and the horizontal velocities derived from the resulting models would be compared with the original model velocities. Repeating this for different noise levels would allow evaluation of the impact of noise on the accuracy of the horizontal velocity determination.

The proposed approach could also be tested by comparison with horizontal velocity fields derived using machine-learning techniques. For example, the DeepVel method \citep{2017A&A...604A..11A} could be used for this purpose.

In the current version of the method, the continuity equation is adopted as the first equation, and the system is closed by an additional constraint of zero vertical vorticity. However, numerical MHD simulations and observations indicate that vortex motions are widespread in the solar atmosphere \citep[see e.g.][]{2023SSRv..219....1T}. This motivates the exploration of alternative closures for obtaining a self-consistent system of equations. Possible options include imposing constraints on the horizontal components of vorticity or employing the Euler equation, either in its full form or via selected components.

An additional and independent constraint on the horizontal velocities can be obtained from the induction equation, provided that the magnetic field vector is known \citep[see e.g.][]{2026A&A...707A.306V}. In this context, comparing the horizontal velocities inferred from the proposed method with those derived from the induction equation would be highly informative. Such a comparison can be performed using realistic MHD simulations, such as the one employed here, and we plan to investigate this in future work.

\section{Conclusions}
\label{sec:conclusions}

We have implemented and tested the method for determining the horizontal velocity field in the solar photosphere, originally proposed by \citet{2016KPCB...32..145S} to reconstruct horizontal flows from inversion results. The approach relies on the continuity equation and the assumption of negligible vertical vorticity. By using a finite-difference approximation for the derivatives, the resulting differential equations are converted into a system of linear algebraic equations. Our improvement of the method consists mainly of an optimised numerical implementation. The sparsity of the coefficient matrix enables both compact storage in memory and fast matrix multiplication, substantially reducing the computational cost of the method.

We applied the method to the 3D MHD Bifrost simulation data to validate its performance and assess intrinsic reconstruction uncertainties. The availability of model horizontal velocities allowed a quantitative comparison between the reconstructed and reference velocity fields.

The current version of the method follows the original approach in employing an iterative procedure for the determination of horizontal velocities. We analysed the dependence of the reconstruction quality on the iteration number at different atmospheric heights and identified the range of iterations providing the best agreement with the model data. This demonstrates that excessive iterations lead to a deterioration of the agreement with the model velocity field, thereby motivating the use of an optimal fixed number of iterations. 

The reconstructed horizontal velocity fields contain artefacts in the form of stripe patterns, attributed to horizontal periodic boundary conditions, as well as localised regions with reduced reconstruction quality. The application of a sinc filter with a Lanczos window effectively suppresses these artefacts, preserving the small-scale structure of the horizontal velocity field. The optimal cutoff frequency was determined independently for each height.

We assessed the method's performance using correlation analysis and investigated the impact of the underlying assumption of zero vertical vorticity on the reliability of the reconstructed velocity fields over a wide height range. Tests using a sequence of simulation snapshots demonstrated that the method provides reliable horizontal velocity estimates in photospheric layers, with the Pearson correlation coefficient between the reconstructed and model velocities reaching 0.8--0.9. The reconstruction is more accurate in granules, while the largest discrepancies are found in intergranular lanes, where complex counter-streaming flows and enhanced vertical vorticity are present. In the chromospheric layers, the reconstruction quality deteriorates significantly, consistent with the increased relative contribution of vortex motions and the resulting breakdown of the zero vertical vorticity assumption, thereby limiting the method's applicability.

The method effectively captures the dominant spatial structures of the velocity field, making it capable of characterising solar atmospheric flows. Its accuracy at granular scales enables application to high-resolution solar observations, while our improvements make the approach better suited for analysing extended fields of view. Overall, the results demonstrate that the improved method offers a reliable and efficient tool for quantitative studies of photospheric dynamics, and they also indicate the limitations in higher atmospheric layers where the underlying assumption is less valid.

\begin{acknowledgements}
	The authors are grateful to the anonymous Referee for valuable comments and suggestions, and especially for the idea of reformulating the methodology in the Fourier domain. The publication is part of the project \emph{``Investigation of the formation of observable characteristics of multicomponent systems in the Universe: cosmic plasma, near-Earth space, and the Earth's atmosphere''} (state registration number 0126U002164), funded by the Ministry of Education and Science of Ukraine.
	
	The authors express their sincere gratitude to all those defending the freedom, independence, and territorial integrity of Ukraine, whose courage and self-sacrifice allow Ukrainian scientists to continue their research.
\end{acknowledgements}

\bibliographystyle{aa}
\bibliography{HorVel}

\begin{appendix}
	\onecolumn
	\section{Schematic representation of matrices $\mathbf{A}$ and $\mathbf{A}^{\rm\!T}\!\mathbf{A}$}
		\begin{figure*}[h!]
			\centering
			\includegraphics[width=17cm]{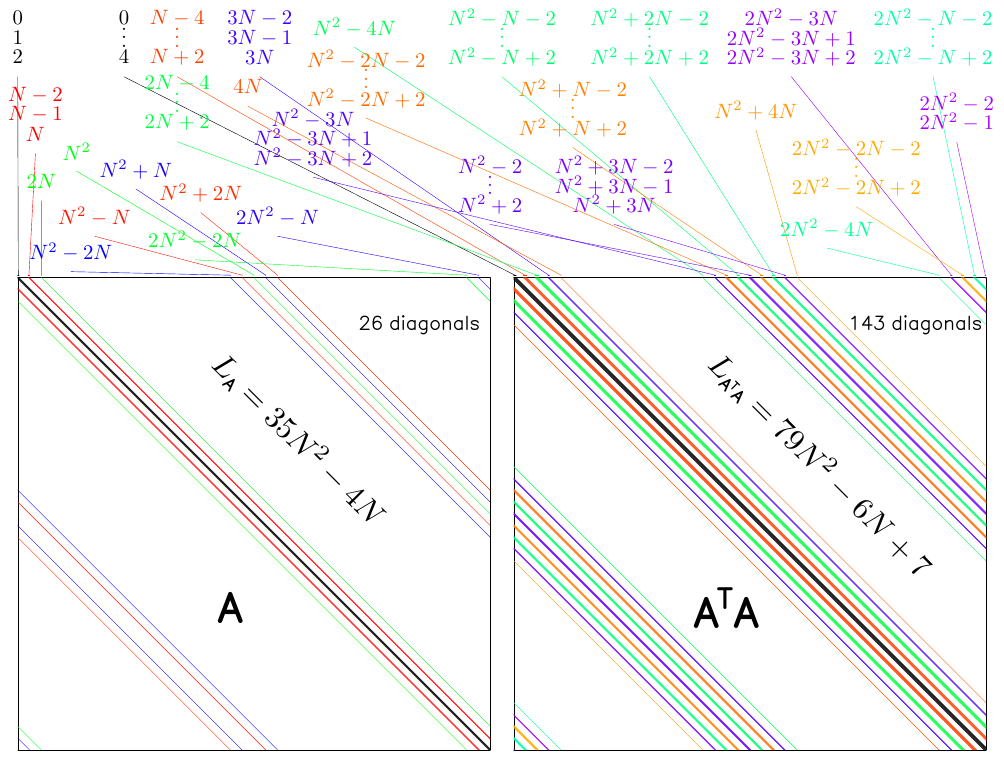}
			\caption{Matrices $\mathbf{A}$ and $\mathbf{A}^{\rm\!T}\!\mathbf{A}$. Coloured lines indicate all diagonals containing non-zero elements. Each colour corresponds to a separate group of diagonals above the main diagonal and to the corresponding group below it. The diagonal displacements are indicated above the matrix diagrams. The relative positions of the diagonals correspond to the case $N=20$. The diagrams show the number of diagonals containing non-zero elements in each matrix, as well as the formulas for their total lengths ($L_\mathbf{A}$ and $L_{\mathbf{A}^{\rm\!T}\!\mathbf{A}}$).}
			\label{fig:matrices}
		\end{figure*}
		\FloatBarrier
		\clearpage
	\section{Examples of filtering results}
		\begin{figure*}[h!]
			\centering
			\includegraphics[width=17cm]{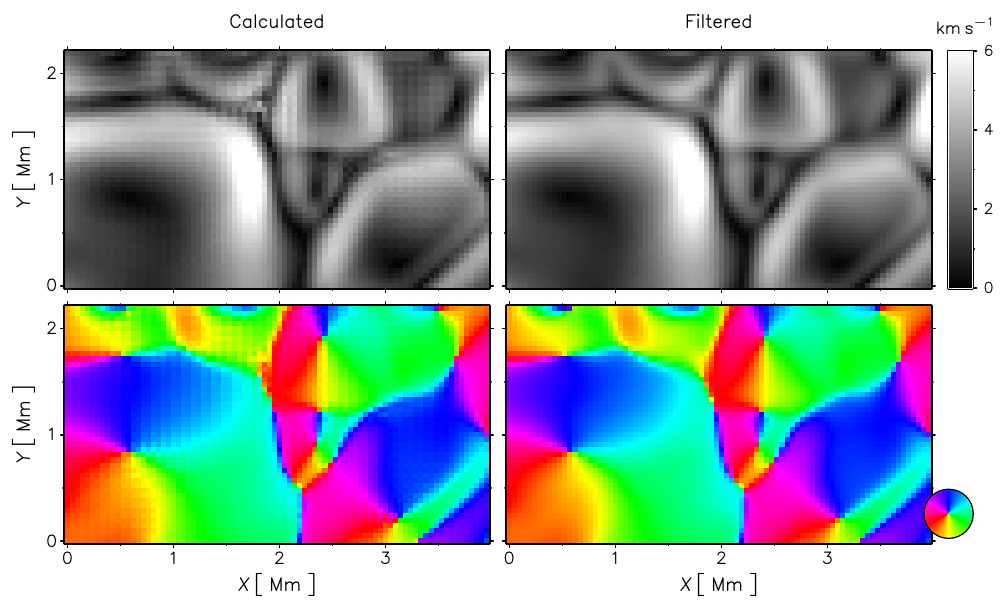}
			\caption{Example of the calculated horizontal velocity field at a height of 160~km (snapshot 391) for selected region exhibiting pronounced artefacts (\textit{left panel}) and the same velocity field after filtering (\textit{right panel}). As in Fig.~\ref{fig:originalandfiltered}, the top panels display the horizontal velocity magnitudes, while the bottom panels indicate flow directions.}
			\label{fig:artefacts1}
		\end{figure*}

		\begin{figure*}[h!]
			\centering
			\includegraphics[width=17cm]{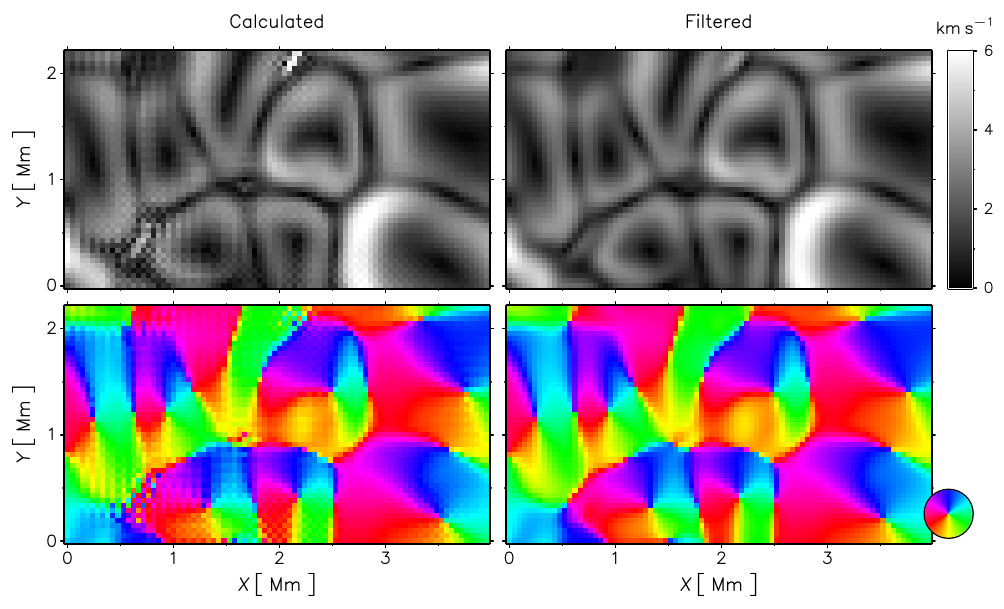}
			\caption{Same as Fig.~\ref{fig:artefacts1} but for a different region.}
			\label{fig:artefacts2}
		\end{figure*}
		\FloatBarrier
		\clearpage
		\twocolumn
	\section{Reformulating the problem in the Fourier domain}
	\label{app:Fourier}
		Since the horizontal velocity field is assumed to be irrotational, a scalar velocity potential $\phi$ can be introduced such that
		\begin{equation*}
			u_x=\frac{\partial\phi}{\partial x},\quad u_y=\frac{\partial\phi}{\partial y}.
		\end{equation*}
		Therefore, the continuity equation can be rewritten as a differential equation for the single unknown function $\phi$:
		\begin{equation*}
			\frac{\partial}{\partial x}\left(\varrho\frac{\partial\phi}{\partial x}\right)+\frac{\partial}{\partial y}\left(\varrho\frac{\partial\phi}{\partial y}\right)=S,
		\end{equation*}
		where
		\begin{equation*}
			S=-\frac{\partial\varrho}{\partial t}-\frac{\partial(\varrho u_z)}{\partial z}.
		\end{equation*}
		
		The density $\varrho$ can be decomposed into its mean value $\bar{\varrho}$ and the perturbation $\varrho-\bar{\varrho}$, which is assumed to be small. The resulting equation is then solved for $\phi$ using an iterative procedure with the initial guess $\phi^{(0)}=0$.
		
		Thus, the equation to be solved is
		\begin{equation}
			\bar{\varrho}\left(\frac{\partial^2\phi^{(n)}}{\partial x^2}+\frac{\partial^2\phi^{(n)}}{\partial y^2}\right)=R^{(n-1)},
			\label{eq:equation}
		\end{equation}
		where the right-hand side $R^{(n-1)}$ is evaluated from the solution $\phi$ obtained at the previous $(n-1)$-th iteration:
		\begin{equation*}
			R^{(n-1)}=S-\frac{\partial}{\partial x}\left([\varrho-\bar{\varrho}]\frac{\partial\phi^{(n-1)}}{\partial x}\right)-\frac{\partial}{\partial y}\left([\varrho-\bar{\varrho}]\frac{\partial\phi^{(n-1)}}{\partial y}\right),\quad R^{(0)}=S.
		\end{equation*}
		
		The derivatives can be evaluated using a 2D DFT on an $N\times N$ grid:
		\begin{equation*}
			\mathcal{F}[\phi](p,q)=\sum_{i=0}^{N-1}\sum_{j=0}^{N-1}\phi(i,j)e^{-\frac{2\pi\textrm{i}}{N}(pi+qj)},
		\end{equation*}
		\begin{equation*}
			\phi(i,j)=\frac{1}{N^2}\sum_{p=0}^{N-1}\sum_{q=0}^{N-1}\mathcal{F}[\phi](p,q)e^{\frac{2\pi\textrm{i}}{N}(pi+qj)}.
		\end{equation*}
		The first and second spectral derivatives with respect to $x$ are expressed as:
		\begin{align}
			\frac{\partial\phi}{\partial x}&=\mathcal{F}^{-1}[\textrm{i}k_p\mathcal{F}[\phi]],\label{eq:firstspectralderivative}\\
			\frac{\partial^2\phi}{\partial x^2}&=-\mathcal{F}^{-1}[k_p^2\mathcal{F}[\phi]],
			\label{eq:secondspectralderivative}
		\end{align}
		where the discrete wavenumber is defined as:
		\begin{equation*}
			k_p=\begin{cases}
				\frac{2\pi p}{L} & 0\le p\le N/2 \\
				\frac{2\pi(p-N)}{L} & N/2<p<N
			\end{cases}.
		\end{equation*}
		Here, $L$ is the horizontal size of the computational domain. Analogous expressions are obtained for the derivatives with respect to $y$ and for the wave number $k_q$.
		
		Since the 2D DFT treats the input as a periodic signal, mismatches at the boundaries introduce high-frequency edge artefacts that manifest as striping patterns. To suppress these artefacts, an apodisation window (e.g. a Tukey window) can be applied to the input data before performing the forward DFT. If the apodisation function contains adjustable parameters, it is necessary to find their optimal values.
		
		Equation~\eqref{eq:equation} is therefore solved for the velocity potential $\phi$ by applying Eq.~\eqref{eq:secondspectralderivative}:
		\begin{equation*}
			\phi^{(n)}=-\mathcal{F}^{-1}\left[\frac{\mathcal{F}\big[R^{(n-1)}\big]}{\bar{\varrho}(k_p^2+k_q^2)}\right].
		\end{equation*}
		The zero-frequency mode is set to zero because the velocity potential is defined only up to an arbitrary additive constant.
		
		\begin{figure*}[h!]
			\centering
			\includegraphics[width=17.0cm]{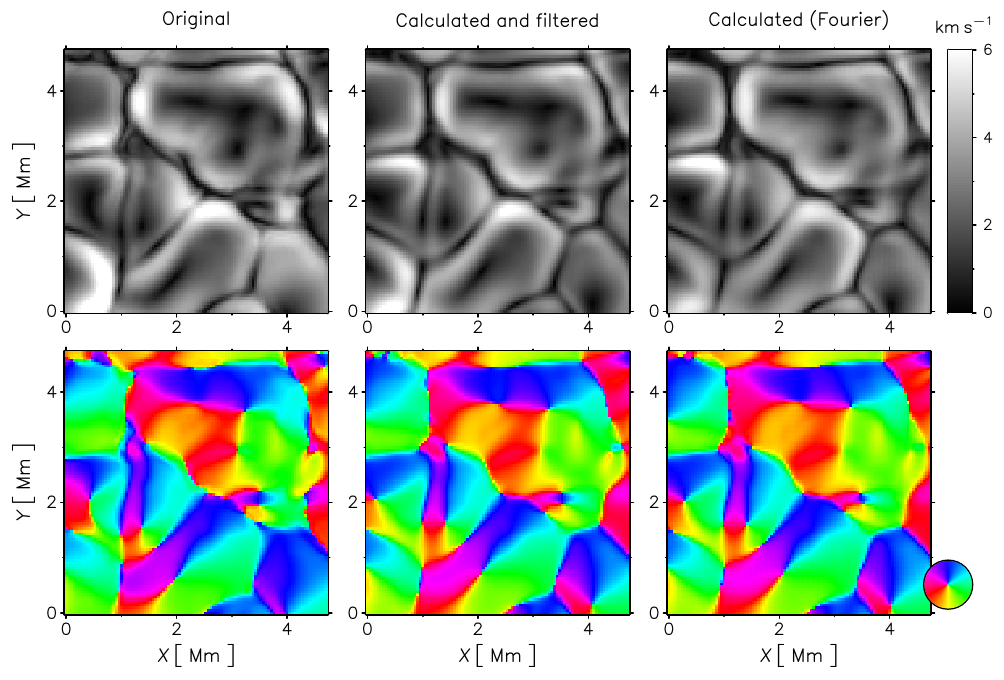}
			\caption{Comparison of horizontal velocity fields obtained using different approaches. \textit{Left and middle panels}: Same data as shown in Fig.~\ref{fig:originalandfiltered} but for the central ($100\times100$ pixels) part of the analysed area. \textit{Right panels}: Corresponding horizontal velocities calculated using the Fourier approach. As in previous figures, the top panels show the absolute values of the horizontal velocity, while the bottom panels show the velocity directions using a colour representation.}
			\label{fig:Fourier}
		\end{figure*}
		
		We use the value of $\phi$ obtained after completing the iterative procedure to compute the horizontal velocity components using Eq.~\eqref{eq:firstspectralderivative} and the analogous expression for the $y$-derivative. The velocity field calculated using the Fourier approach for snapshot 391 at a height of 160~km is shown in Fig.~\ref{fig:Fourier}, together with the corresponding model data and the velocity field calculated using the method described in this paper. Comparison with the method presented in the main text indicates that the Fourier approach substantially reduces the computational time while maintaining a comparable reconstruction quality.
		
		\FloatBarrier
\end{appendix}

\end{document}